\documentclass[aps, notitlepage, floatfix, pre, twocolumn]{revtex4-2}
\usepackage[colorlinks=true, allcolors=blue, bookmarks=false]{hyperref}
\usepackage{amsfonts, amsmath, amssymb}
\usepackage{graphicx}

\setcitestyle{numbers,square,sort&compress}
\DeclareMathOperator{\Tr}{Tr}

\begin{document}

\title{Solvable random matrix ensemble with a logarithmic weakly confining potential}

\author{Wouter Buijsman} 

\email{buijsman@post.bgu.ac.il}

\affiliation{Department of Physics, Ben-Gurion University of the Negev, Beer-Sheva 84105, Israel}

\date{\today}

\begin{abstract}
This work identifies a solvable (in the sense that spectral correlation functions can be expressed in terms of orthogonal polynomials), rotationally invariant random matrix ensemble with a logarithmic weakly confining potential. The ensemble, which can be interpreted as a transformed Jacobi ensemble, is in the thermodynamic limit characterized by a Lorentzian eigenvalue density. It is shown that spectral correlation functions can be expressed in terms of the nonclassical Gegenbauer polynomials $C_n^{(-1/2)}(x)$ with $n \ge 2$, which have been proven to form a complete orthogonal set with respect to the proper weight function. A procedure to sample matrices from the ensemble is outlined and used to provide a numerical verification for some of the analytical results. This ensemble is pointed out to potentially have applications in quantum many-body physics.
\end{abstract}

\maketitle

%%%%%%%%%%%%%%%%%%%%%%%%%%%%%%
\section{Introduction}
%%%%%%%%%%%%%%%%%%%%%%%%%%%%%%
Random matrix theory plays a major role in the analysis of various types of complex quantum systems \cite{Guhr98, Mehta10} with applications, for example, in nuclear physics \cite{Weidemuller09, Borgonovi16}, mesoscopic physics \cite{Beenakker97, Beenakker15}, high-energy physics \cite{Maldacena16, Cotler18}, and quantum chaos \cite{DAlessio16, Deutsch18}. One of the main challenges in physically-motivated random matrix theory is to construct random matrix models that are on the one hand simple enough to be tractable analytically, and on the other hand, provide a reasonably good description of the system of interest. Notable progress in the search for such ensembles has been made in the last decade \cite{Kravtsov15, vonKeyserlingk18, Nahum18, Chan18}. 

The central building blocks of random matrix theory are the three classical (Gaussian, Wishart-Laguerre, and Jacobi) random matrix ensembles (see, e.g., Refs. \cite{Forrester10, Livan18}). These ensembles are rotationally invariant (i.e., basis-independent) and solvable in the sense that spectral correlation functions can be expressed in terms of orthogonal polynomials. For these ensembles, the joint probability distribution for the eigenvalues is known explicitly. Expressing this distribution as the Boltzmann factor of a Coulomb gas trapped in a confining potential allows one to study the thermodynamic limit using tools from statistical mechanics.

This work identifies a solvable, rotationally invariant random matrix ensemble with a logarithmic (weakly) confining potential. In the thermodynamic limit, the eigenvalue densities of the Gaussian, Wishart-Laguerre, and Jacobi ensembles are given by respectively the Wigner semicircle, Mar\v{c}enko-Pastur, and Wachter laws \cite{Wachter78}. Using the Coulomb gas technique, the eigenvalue density in the thermodynamic limit corresponding to the logarithmic potential is found to be given by a Lorentzian. Random matrices with a Lorentzian eigenvalue density appeared very recently in the context of ergodicity breaking in quantum many-body systems in Ref. \cite{Venturelli22}. 

The orthogonal polynomials in terms of which spectral correlation functions can be expressed are identified as the nonclassical Gegenbauer polynomials $C_n^{(-1/2)}(x)$ with $n \ge 2$, which have been proven to form a complete orthogonal set with respect to the proper weight function \cite{Bruder14}. The Gegenbauer polynomials form a subset of the Jacobi polynomials. From this, it is deduced that the ensemble can be interpreted as a transformed Jacobi ensemble. A procedure to numerically sample eigenvalue spectra from the ensemble is outlined and demonstrated by verifying some of the analytical results.

The outline of this work is as follows. Section \ref{sec: Coulomb} considers the Coulomb gas picture for the joint probability distribution of the eigenvalues. Here, it is discussed how the Lorentzian density of states emerges from the logarithmic potential. Section \ref{sec: polynomials} identifies the associated (nonclassical) orthogonal polynomials, and outlines how the spectral correlation functions can be obtained. Here, the relation with the Jacobi ensemble is also discussed. Section \ref{sec: matrix} outlines how the eigenvalue spectra can be obtained from the spectra of random matrices. Section \ref{sec: conclusions} provides a summary of the findings and proposes suggestions for further investigations.

%%%%%%%%%%%%%%%%%%%%%%%%%%%%%%
\section{Coulomb gas picture} \label{sec: Coulomb}
%%%%%%%%%%%%%%%%%%%%%%%%%%%%%%
For the classical random matrix ensembles, the joint probability distribution $P(H)$ of the entries of sampled matrices $H$ can be written as
\begin{equation}
P(H) \varpropto \exp[ -\Tr V(H)],
\end{equation} 
where $V(x)$ is a function referred to as the potential (see, e.g., chapters 4 and 5 of Ref. \cite{Livan18}). Let $N$ denote the dimension of the matrices. As $P(H)$ depends only on (powers of) the trace of $H$, the ensembles are rotationally invariant. That is, the ensembles are invariant under transformations of the basis. The joint probability distribution $P(x_1, \dots, x_N)$ of the eigenvalues $x_n$ is in terms of the potential given by
\begin{align}
& P(x_1, \dots, x_N) = \frac{1}{\mathcal{Z}_{N \beta}} e^{- \beta \, \mathcal{V}(x_1, \dots, x_N)}, \label{eq: Px} \\
& \mathcal{V} = \frac{1}{\beta} \sum_{n = 1}^N V(x_n) + \frac{1}{2} \sum_{n, m = 1}^N \ln | x_n - x_m|. \label{eq: pot}
\end{align}
Here, $\beta \in \{1, 2, 4\}$ is the Dyson index giving the number of degrees of freedom per (real, imaginary, or quaternionic) matrix element. Next, $\mathcal{Z}_{N \beta}$ is a normalization constant fixing the integrated probability to unity.

Equation \eqref{eq: Px} can be thought of as the Boltzmann factor of a one-dimensional gas of particles interacting through a pairwise logarithmic potential, confined by the one-body potential. Such a gas is commonly referred to as a Coulomb gas \cite{Forrester10}. This interpretation allows one to use tools from statistical mechanics to study the thermodynamic limit. After making a continuum and saddle-point approximation (see, e.g., chapter 5 of Ref. \cite{Livan18}), the eigenvalue density $\rho_N(x)$ for a potential $V(x)$ can be shown to satisfy the integral equation
\begin{equation}
\Pr \int_{- \infty}^\infty \frac{\rho_N(y)}{x - y} dy = \frac{1}{\beta} \frac{dV}{dx},
\label{eq: V-rho}
\end{equation}
where $\Pr$ denotes the prinipal value. This integral equation is subject to the constraint $\int \rho_N(x) \, dx = N$. Equation \eqref{eq: V-rho} is generically difficult to solve, and only for a limited number of potentials the corresponding eigenvalue density has been found (see, e.g., Sec. 3.2 of Ref. \cite{Sakhnovich15}). 

In this work, the focus is on the random matrix ensemble associated with the logarithmic potential
\begin{equation}
V(x) = \frac{\beta N}{2} \ln(1 + x^2).
\label{eq: V}
\end{equation}
For $|x| \gg 1$, this potential approximates $\beta N \ln(x)$. The prefactor $N$ ensures that the first and second terms in Eq. \eqref{eq: pot} are of the same order, namely $N^2$. In the absence of this prefactor, the first term in Eq. \eqref{eq: pot} would be vanishingly small compared to the second one, making the potential nonconfining. The term ``weakly confining'' appeared in the current context first in Ref. \cite{Hardy12}, which discusses the potential studied in this work in Example 1.3.

At a technical level, the motivation to consider this particular potential is as follows. Substituting Eq. \eqref{eq: V} in Eq. \eqref{eq: V-rho} and dividing both sides by $N$ gives on the left-hand side the Hilbert transform (see, e.g., chapter 5 of Ref. \cite{Liflyand21}) $\mathcal{H}[f(y)]$ of some function $f(y)$,
\begin{equation}
\mathcal{H}[f(y)] = \frac{1}{\pi} \Pr \int_{- \infty}^\infty \frac{f(y)}{x - y} dy.
\end{equation}
By comparing the right-hand side $x / (1 + x^2)$ with known Hilbert transforms, one deduces that the eigenvalue density is given by
\begin{equation}
\lim_{N \to \infty} \rho_N(x) = \frac{N}{\pi (1 + x^2)},
\end{equation}
which is referred to as $\rho(x)$ below. An explicit derivation can be found, e.g,. in example 5.17 of the reference cited above. Indeed, this function can easily be shown to obey the normalization condition $\int \rho_N(x) \, dx = N$. This density sharply differs from, e.g., the semicircular eigenvalue density as observed for the Gaussian ensembles.

It can be of interest to note that random matrix ensembles with logarithmic or squared-logarithmic potentials (although without prefactor $N$) have been proposed as models for the intermediate level spacing statistics and multifractality at the Anderson localization transition \cite{Akemann15} (chapter 12), \cite{Muttalib93, Evers08, Choi10, Vleeshouwers21}.

%%%%%%%%%%%%%%%%%%%%%%%%%%%%%%
\section{Orthogonal polynomials} \label{sec: polynomials}
%%%%%%%%%%%%%%%%%%%%%%%%%%%%%%
Spectral correlation functions for the classical random matrix ensembles at finite dimension can be expressed in terms of orthogonal polynomials (see, e.g., chapter 10 of Ref. \cite{Livan18}). In view of the discussion below, the main ideas are introduced using the Jacobi ensemble as an illustration. The Jacobi ensemble is known historically to be relevant in physics in the context of quantum conductance \cite{Forrester06}. In recent years, new applications appeared in the computation of eigenstate entanglement of random free fermionic models \cite{Liu18, Lydzba21, Bianchi21, Murciano22, Ulcakar22} and the spectral form factor of the self-dual kicked Ising model \cite{Flack20}.

As before, let $N$ and $\beta$ denote the dimension of the matrices and the Dyson index, respectively. The eigenvalues $x_n \in [-1,1]$ of samples from the Jacobi ensemble are distributed according to
\begin{equation}
P(x_1, \dots, x_N) = \frac{1}{\mathcal{Z}_{a b N \beta}} \prod_{n = 1}^N w(x_n) \prod_{m < k} |x_m - x_k|^\beta
\label{eq: P-Jacobi}
\end{equation}
with the weight function $w(x)$ characterized by parameters $a > -1$ and $b > -1$ here given by
\begin{equation}
w(x) = (1 - x)^{a \beta / 2} \, (1 + x)^{b \beta / 2}.
\end{equation}
Similar to the above, $\mathcal{Z}_{a b N \beta}$ is a normalization constant fixing the integrated probability to unity. For probability distributions of the form \eqref{eq: P-Jacobi}, spectral correlation functions can be expressed for $\beta = 2$ in terms of the kernel 
\begin{equation}
K_N(x_1, x_2) = e^{- \frac{1}{2} \big( V(x_1) + V(x_2) \big)} \sum_{n = 0}^{N-1} p_n (x_1) \, p_n (x_2),
\label{eq: K}
\end{equation}
where, for notational convenience, the potential $V(x)$ satisfying $w(x) = e^{-V(x)}$ is reintroduced. The functions $p_n(x)$ are polynomials orthogonal with respect to the weight function. For the Jacobi ensemble, thus
\begin{equation}
\int_{-1}^1 (1 - x)^{a \beta / 2} (1 + x)^{b \beta / 2} \, p_n(x) \, p_m(x) \, dx = \delta_{nm}.
\label{eq: Jacobi-orthogonality}
\end{equation}
For Eq. \eqref{eq: Jacobi-orthogonality}, the polynomials $p_n(x)$ are given by the Jacobi polynomials $P_n^{(a,b)}(x)$ (up to normalization) with the same parameters (see, e.g., Sec. 9.8 of Ref. \cite{Koekoek10} or chapter 4 of Ref. \cite{Szego75}). In terms of the kernel, the eigenvalue density $\rho_N (x)$ is given by
\begin{equation}
\rho_N(x) = K_N(x, x).
\label{eq: rho1}
\end{equation}
Two-point eigenvalue correlation functions can be expressed in terms of the kernel as
\begin{equation}
\rho_N^{(2)}(x_1, x_2) = C_N \det
\begin{pmatrix}
K_N(x_1, x_1) & K_N(x_1, x_2) \\
K_N(x_2, x_2) & K_N(x_2, x_2)
\end{pmatrix}
\label{eq: rho2}
\end{equation}
with $C_N = 1 / [N (N-1)]$.

Eigenvalue correlation functions for the potential of Eq. \eqref{eq: V} can be studied by first making the change of variables $x \to y$ given by
\begin{equation}
y = \frac{x}{\sqrt{1 + x^2}}.
\label{eq: transformation}
\end{equation}
From the inverse relation $x = y / \sqrt{1 - y^2}$, it follows that the orthogonality condition for the potential studied in this work is in terms of $y$ given by
\begin{equation}
\int_{-1}^1 \frac{1}{1 - y^2} \, p_n(y) \, p_m(y) = \delta_{nm}.
\end{equation}
One recognizes the orthogonality condition for the nonclassical Gegenbauer polynomials $C_n^{(\lambda)}(y)$, which are (up to a prefactor) Jacobi polynomials $P_n^{(a, b)}(y)$ with $a = b = \lambda - 1/2$ [Eq. \eqref{eq: Jacobi-orthogonality}], with $\lambda = -1/2$. See the Appendix for details.

For $\lambda = -1/2$, the first two Gegenbauer polynomials ($n = 0$ and $n = 1$) are not normalizable (hence the classification ``nonclassical''). As mentioned above, the Gegenbauer polynomials $\{ C_n^{(-1/2)}(y) \}_{n=2}^\infty$ are known to form a complete orthogonal set with respect to the proper weight function \cite{Bruder14}. In the evaluation of the kernel [Eq. \eqref{eq: K}], the counting thus starts at $n=2$, due to which the summation runs up to $n = N+1$. 

For the random matrix ensemble studied in this work, eigenvalue correlations can thus be obtained in terms of the variable $y$. The relation $x = y / \sqrt{1 - y^2}$ allows one to subsequently obtain correlations in terms of the original variable $x$. Having found the orthogonal polynomials, the ensemble proposed in this work can be considered as being ``solvable'' (see, e.g., Ref. \cite{Muttalib93} for details on this classification).

\begin{figure}[t]
\includegraphics[scale=0.9]{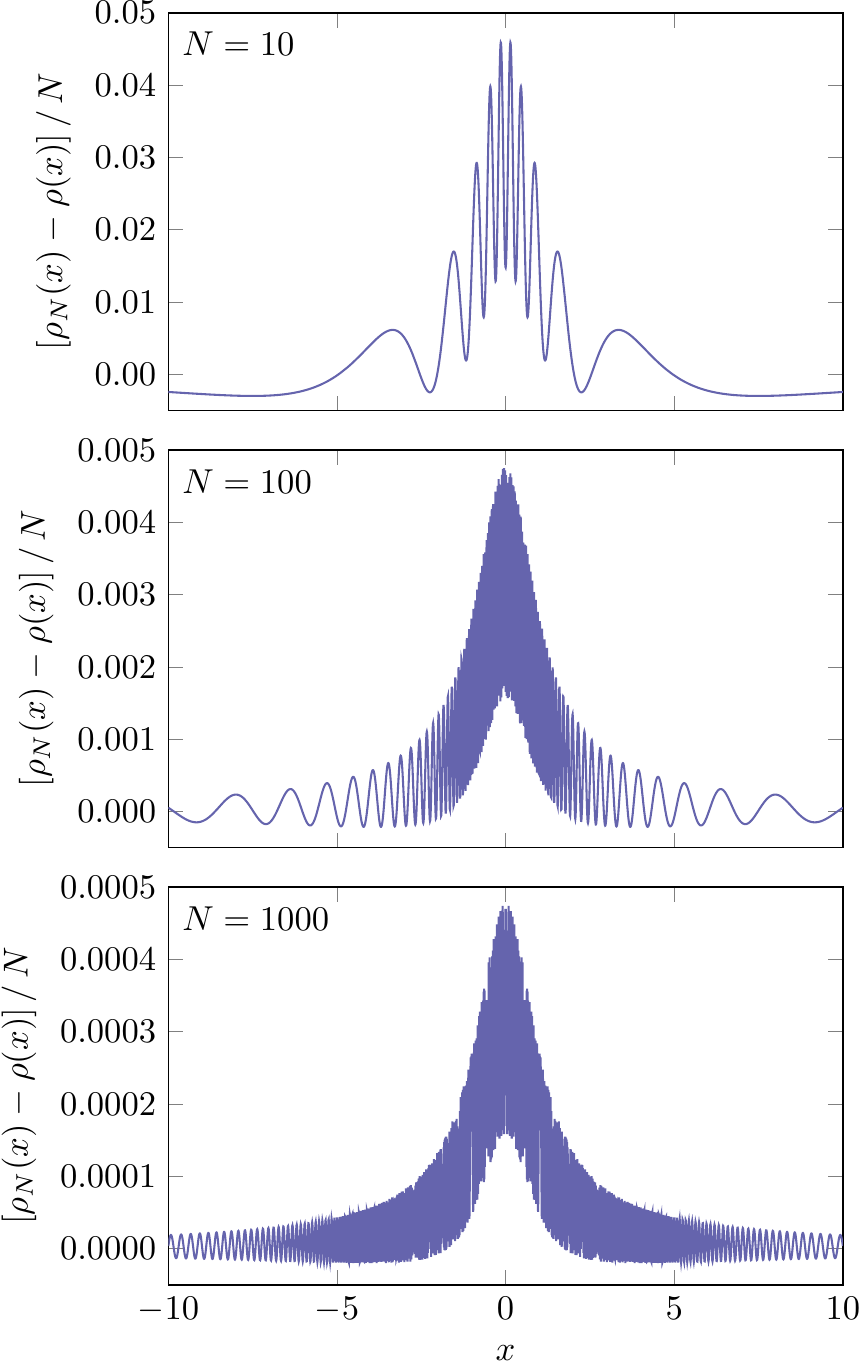}
\caption{The difference between the normalized (to unity) eigenvalue density for $N$ finite and $N \to \infty$ for $N = 10$, $N = 100$, and $N = 1000$. One observes that the difference scales as $1/N$.}
\label{fig: rho1}
\end{figure}

Aiming to illustrate the above results, here some numerical evaluations of $\rho_N(x)$ [Eq. \eqref{eq: rho1}] and $\rho_N^{(2)}(0,x)$ [Eq. \eqref{eq: rho2}] are presented. Figure \ref{fig: rho1} shows the difference between the normalized (to unity) eigenvalue density for $N$ finite and $N \to \infty$ for $N = 10$, $N = 100$, and $N = 1000$. The difference becomes smaller with increasing $N$, scaling as $1/N$. The data for each of the figures presented in this work (except for the histograms in Fig. \ref{fig: hist}) can be generated in $\sim \hspace{-1.5mm} 10$ minutes of computational time on a midrange laptop.

\begin{figure}[t]
\includegraphics[scale=0.9]{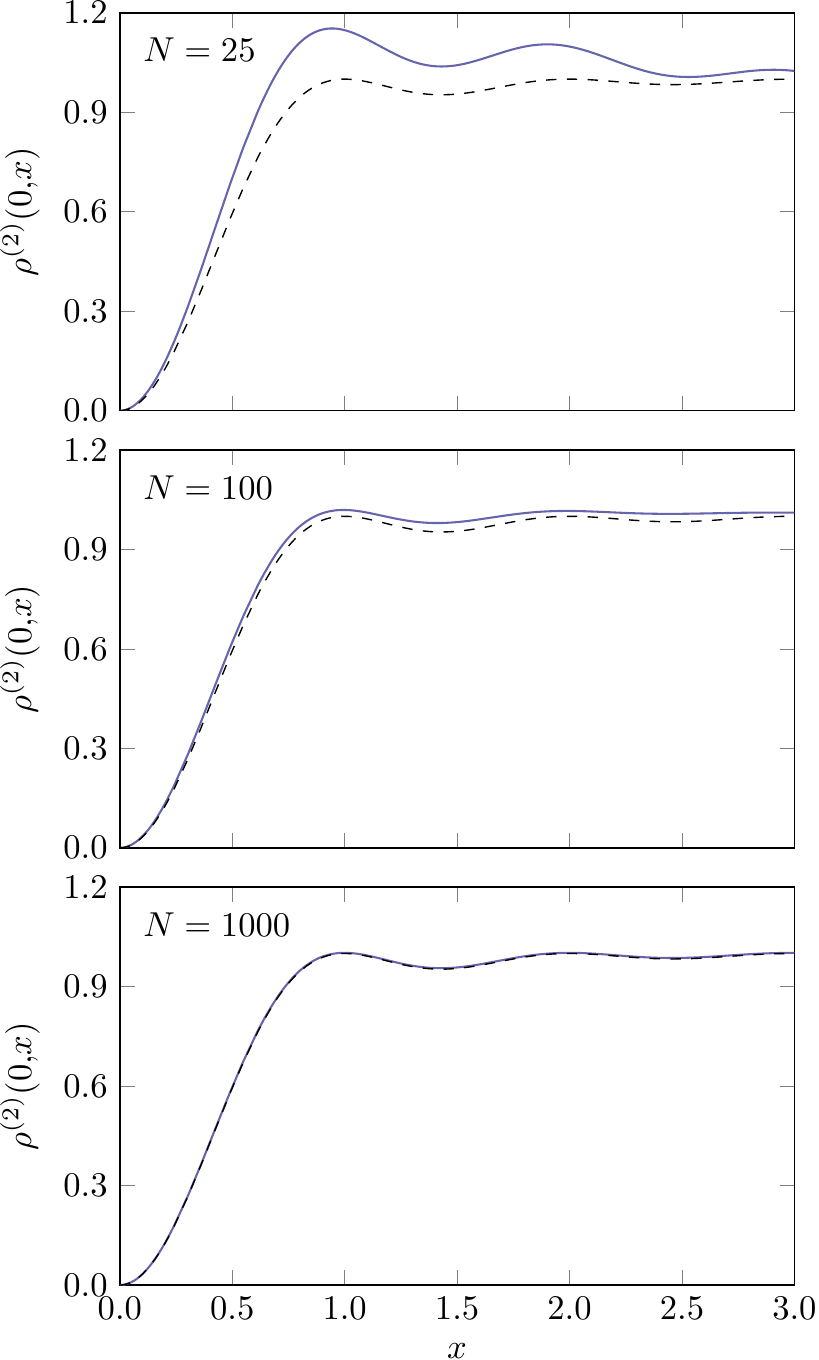}
\caption{A comparison between $\rho_N^{(2)}(0,x)$ (solid lines) and the result for Wigner-Dyson level statistics at $N \to \infty$ of Eq. \eqref{eq: RMT-rho2} (dashed lines) for $N = 25$, $N = 100$, and $N = 1000$. The finite-$N$ results have been scaled by $1 / [\rho(0) / N]^2 = \pi^2$ and the linear transformation $x \to \pi x / N$ has been applied in order to set the mean level spacing at $x = 0$ to unity for $N \to \infty$. The finite-$N$ curves approach the Wigner-Dyson result as $N$ increases.}
\label{fig: rho2}
\end{figure}

Figure \ref{fig: rho2} compares $\rho^{(2)}_N(0,x)$ for $N = 25$, $N = 100$, and $N = 1000$ with the evaluation for Wigner-Dyson level statistics at $N \to \infty$ for unfolded spectra with unit mean level spacing given by
\begin{equation}
\rho^2(0,x) = 1 - \bigg( \frac{\sin(\pi x)}{\pi x} \bigg)^2,
\label{eq: RMT-rho2}
\end{equation}
see e.g. Ref. \cite{Mehta10}. The finite-$N$ results have been scaled and transformed such that the mean level spacing is unity at $x = 0$ for $N \to \infty$ (see the caption for details). The finite-$N$ curves approach the Wigner-Dyson result as $N$ increases. For $N = 25$, effects due to the non-uniform eigenvalue density (decaying with increasing $x$) are clearly visible. For $N = 1000$, the curves are visually indistinguishable.

%%%%%%%%%%%%%%%%%%%%%%%%%%%%%%
\section{Random matrix construction} \label{sec: matrix}
%%%%%%%%%%%%%%%%%%%%%%%%%%%%%%
In Sec. \ref{sec: polynomials}, it was found that the spectral correlation functions for the random matrix ensemble proposed in this work can be obtained from the spectral correlation functions for the Jacobi ensemble with parameters $a = b = -1$ through the transformation given in Eq. \eqref{eq: transformation}. This mapping of the spectral properties of a classical ensemble to the spectral properties of the ensemble of interest allows one to sample spectra by diagonalizing random matrices.

Let $X_1$ and $X_2$ denote, respectively, $M_1 \times N$ and $M_2 \times N$ matrices with independent sampled Gaussian entries. The real ($\beta = 1$), imaginary ($\beta = 2$), or quaternionic-valued ($\beta = 4$) entries $x$, $z$, or $w$ are sampled from, respectively, the probability densities
\begin{equation}
\frac{1}{\sqrt{2 \pi}} e^{- \frac{1}{2} x^2}, \qquad \frac{1}{\pi} e^{-|z|^2}, \qquad \frac{2}{\pi} e^{-2 |w|^2},
\end{equation}
see, e.g., Sec. 6.3\ of Ref. \cite{Forrester10}. Next, let $W_1 = X_1^\dagger X_1$ and $W_2 = X_2^\dagger X_2$. A spectrum $\{ y_n \}$ from the Jacobi ensemble of dimension $N$ with parameters $a = N - M_1 + 1 - 2/\beta$ and $b = N - M_2 + 1 - 2/\beta$ is obtained by transforming the eigenvalues $\{ z_n \}$ of the double-Wishart matrix
\begin{equation}
W = W_1 (W_1 + W_2)^{-1},
\label{eq: W-matrix}
\end{equation}
which obey $z_n \in [0,1]$, as $y_n = 1 - 2 z_n$. Notice that there are no issues with sampling for $a = b = -1$, which can be accomplished by choosing $M_1 = M_2 = N + 1$.

\begin{figure}[t]
\includegraphics[scale=0.9]{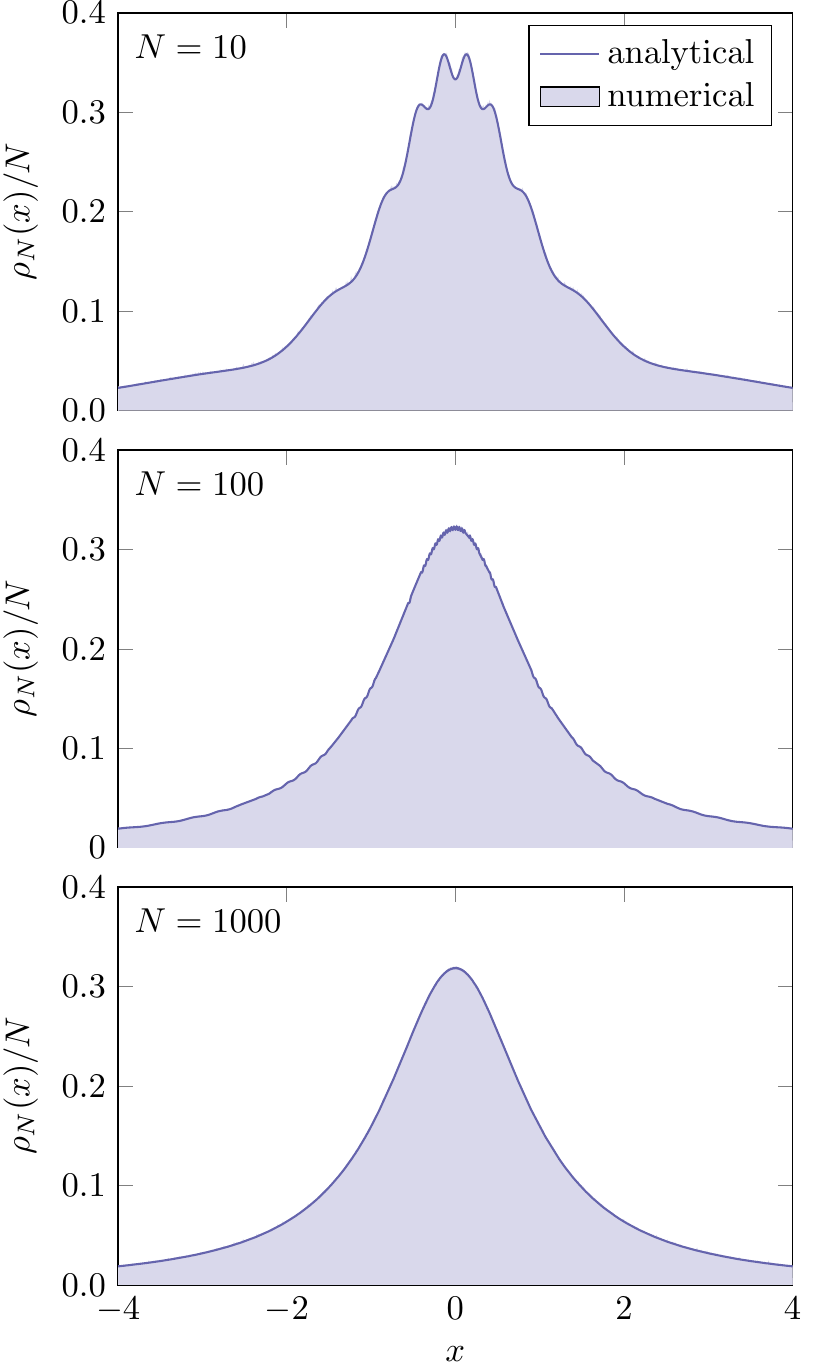}
\caption{A plot of the normalized (to unity) eigenvalue densities $\rho_N(x) / N$ (``analytical'') compared with a normalized [to unity on $x \in (- \infty, \infty)$] histogram of the eigenvalues for the random matrix ensemble proposed in this work (``numerical'') for $\beta = 2$ at $N = 10$, $N = 100$, and $N =1000$. Perfect agreement can be observed.}
\label{fig: hist}
\end{figure}

Given an eigenvalue spectrum $\{ y_n \}$ sampled from the Jacobi ensemble, a spectrum $\{ x_n \}$ of the ensemble proposed in this work can be obtained by applying the inverse of the transformation $x_n \to y_n$ as given in Eq. \eqref{eq: transformation}, $x = y / \sqrt{1 - y^2}$. Figure \ref{fig: hist} compares the normalized (to unity) eigenvalue density $\rho_{N}(x) / N$ with a properly normalized histogram of the eigenvalues for $\beta = 2$ at $N = 10$, $N = 100$, and $N = 1000$ from the ensemble proposed in this work, obtained through diagonalizations of matrices $W$ as given in Eq. \eqref{eq: W-matrix}. One observes perfect agreement.

%%%%%%%%%%%%%%%%%%%%%%%%%%%%%%
\section{Conclusions and outlook} \label{sec: conclusions}
%%%%%%%%%%%%%%%%%%%%%%%%%%%%%%
In this work, a solvable (in the sense that spectral correlation functions can be expressed in terms of orthogonal polynomials), rotationally invariant random matrix ensemble with a logarithmic weakly confining potential has been identified. This ensemble is found to be a transformed Jacobi ensemble. Using the Coulomb gas technique, the eigenvalue density in the thermodynamic limit is found to be given by a Lorentzian. A procedure to sample numerically from this random matrix ensemble has been outlined, and used to verify some of the analytical results.

As the random matrix ensemble identified in this work can be interpreted as a transformed Jacobi ensemble, properties of the ensemble that have not been discussed here, such as extreme value statistics \cite{MorenoPozas19} or the extension to a continous $\beta$-ensemble \cite{Dumitriu02}, could in principle be established in a straightforward way. It would be of interest to see how the ensemble proposed in this work appears in physical settings. For example, in the spirit of Ref. \cite{Venturelli22} and other generalizations \cite{Khaymovich20, Biroli21, Khaymovich21, Buijsman22, DeTomasi22}, this ensemble could potentially serve as a building block for improved generalizations of the Rosenzweig-Porter ensemble \cite{Kravtsov15}.

%%%%%%%%%%%%%%%%%%%%%%%%%%%%
% Acknowledgments
%%%%%%%%%%%%%%%%%%%%%%%%%%%%
\begin{acknowledgments}
Useful discussions with Vladimir Gritsev and Ward L. Vleeshouwers are gratefully acknowledged. This work is supported by the Kreitman School of Advanced Graduate Studies at Ben-Gurion University.
\end{acknowledgments}

\appendix*
%%%%%%%%%%%%%%%%%%%%%%%%%%%%%%
\section{Gegenbauer polynomials} \label{app: Gegenbauer}
%%%%%%%%%%%%%%%%%%%%%%%%%%%%%%
The Gegenbauer polynomials $C_n^{(\lambda)}(x)$ are Jacobi polynomials $P_n^{(a, b)}(x)$ with $a = b = \lambda - 1/2$ for $\lambda > -1/2$ (see e.g. Sec. 9.8.1 of Ref. \cite{Koekoek10}). Following customary normalization (see, e.g., Sec. 4.7 of Ref. \cite{Szego75}), these are defined as
\begin{equation}
C_n^{(\lambda)}(x) = \frac{\Gamma(\lambda + 1/2)}{\Gamma(2 \lambda)} \frac{\Gamma(n + 2 \lambda)}{\Gamma(n + \lambda + 1/2)} P_n^{(\lambda - 1/2, \lambda - 1/2)}(x).
\end{equation}
The Gegenbauer polyonomals are orthogonal with respect to the weight function $(1+x^2)^{\lambda - 1/2}$ on $x \in [-1,1]$. They obey the recurrence relation
\begin{equation}
\begin{split}
& 2 (n+\lambda) x \, C_n^{(\lambda)}(x) = (n+1) \, C_{n+1}^{(\lambda)}(x) \\
& + (n + 2 \lambda - 1) \, C_{n-1}^{(\lambda)}(x) 
\end{split}
\end{equation}
with $C_0^{(\lambda)}(x) = 1$ and $C_1^{(\lambda)}(x) = 2 \lambda x$. The normalization condition for the Gegenbauer polynomials reads
\begin{equation}
\int_{-1}^1 (1 - x^2)^{\lambda - 1/2} \big[ C_n^{(\lambda)}(x) \big]^2 dx = \mathcal{N}_n^{(\lambda)}
\end{equation}
with
\begin{equation}
 \mathcal{N}_n^{(\lambda)} = \frac{\pi \Gamma(n + 2 \lambda) 2^{1 - 2 \lambda}}{n! (n + \lambda) [ \Gamma(\lambda)]^2}.
\end{equation}
The polynomials satisfying the orthogonality condition of Eq. \eqref{eq: Jacobi-orthogonality} are given by $p_n(x) = C_n^{(\lambda)} (x) / \sqrt{ \mathcal{N}_n^{(\lambda)}}$. A recent generalization of the orthogonality condition to the complex plane has been obtained in Ref. \cite{Akemann21}.

For $\lambda = -1/2$, $ \mathcal{N}_n^{(\lambda)}$ reduces to $1 / [n (n-1/2)(n-1)]$, meaning that the polynomials with indices $n = 0$ and $n = 1$ can not be properly normalized. The Gegenbauer polynomials $\{ C_n^{(-1/2)}(x) \}_{n=2}^\infty$ have been proven to form a complete orthogonal set with respect to the proper weight function, and are classified as ``nonclassical'' \cite{Bruder14}. Consequently, for $\lambda = -1/2$ the counting starts at $n = 2$.

\nocite{apsrev42Control}
\bibliography{references}

%apsrev4-2.bst 2019-01-14 (MD) hand-edited version of apsrev4-1.bst
%Control: key (0)
%Control: author (0) dotless jnrlst
%Control: editor formatted (1) identically to author
%Control: production of article title (0) allowed
%Control: page (1) range
%Control: year (0) verbatim
%Control: production of eprint (0) enabled
\begin{thebibliography}{44}%
\makeatletter
\providecommand \@ifxundefined [1]{%
 \@ifx{#1\undefined}
}%
\providecommand \@ifnum [1]{%
 \ifnum #1\expandafter \@firstoftwo
 \else \expandafter \@secondoftwo
 \fi
}%
\providecommand \@ifx [1]{%
 \ifx #1\expandafter \@firstoftwo
 \else \expandafter \@secondoftwo
 \fi
}%
\providecommand \natexlab [1]{#1}%
\providecommand \enquote  [1]{``#1''}%
\providecommand \bibnamefont  [1]{#1}%
\providecommand \bibfnamefont [1]{#1}%
\providecommand \citenamefont [1]{#1}%
\providecommand \href@noop [0]{\@secondoftwo}%
\providecommand \href [0]{\begingroup \@sanitize@url \@href}%
\providecommand \@href[1]{\@@startlink{#1}\@@href}%
\providecommand \@@href[1]{\endgroup#1\@@endlink}%
\providecommand \@sanitize@url [0]{\catcode `\\12\catcode `\$12\catcode
  `\&12\catcode `\#12\catcode `\^12\catcode `\_12\catcode `\%12\relax}%
\providecommand \@@startlink[1]{}%
\providecommand \@@endlink[0]{}%
\providecommand \url  [0]{\begingroup\@sanitize@url \@url }%
\providecommand \@url [1]{\endgroup\@href {#1}{\urlprefix }}%
\providecommand \urlprefix  [0]{URL }%
\providecommand \Eprint [0]{\href }%
\providecommand \doibase [0]{https://doi.org/}%
\providecommand \selectlanguage [0]{\@gobble}%
\providecommand \bibinfo  [0]{\@secondoftwo}%
\providecommand \bibfield  [0]{\@secondoftwo}%
\providecommand \translation [1]{[#1]}%
\providecommand \BibitemOpen [0]{}%
\providecommand \bibitemStop [0]{}%
\providecommand \bibitemNoStop [0]{.\EOS\space}%
\providecommand \EOS [0]{\spacefactor3000\relax}%
\providecommand \BibitemShut  [1]{\csname bibitem#1\endcsname}%
\let\auto@bib@innerbib\@empty
%</preamble>
\bibitem [{\citenamefont {Guhr}\ \emph {et~al.}(1998)\citenamefont {Guhr},
  \citenamefont {M\"uller-Groeling},\ and\ \citenamefont
  {Weidenm\"uller}}]{Guhr98}%
  \BibitemOpen
  \bibfield  {author} {\bibinfo {author} {\bibfnamefont {T.}~\bibnamefont
  {Guhr}}, \bibinfo {author} {\bibfnamefont {A.}~\bibnamefont
  {M\"uller-Groeling}},\ and\ \bibinfo {author} {\bibfnamefont {H.~A.}\
  \bibnamefont {Weidenm\"uller}},\ }\bibfield  {title} {\bibinfo {title}
  {Random-matrix theories in quantum physics: common concepts},\ }\href
  {https://doi.org/10.1016/S0370-1573(97)00088-4} {\bibfield  {journal}
  {\bibinfo  {journal} {Phys. Rep.}\ }\textbf {\bibinfo {volume} {299}},\
  \bibinfo {pages} {189} (\bibinfo {year} {1998})}\BibitemShut {NoStop}%
\bibitem [{\citenamefont {Mehta}(2010)}]{Mehta10}%
  \BibitemOpen
  \bibfield  {author} {\bibinfo {author} {\bibfnamefont {M.~L.}\ \bibnamefont
  {Mehta}},\ }\href@noop {} {\emph {\bibinfo {title} {{Random Matrices}}}},\
  \bibinfo {edition} {3rd}\ ed.,\ \bibinfo {series} {Pure and Applied
  Mathematics}, Vol.\ \bibinfo {volume} {142}\ (\bibinfo  {publisher}
  {Elsevier},\ \bibinfo {address} {New York},\ \bibinfo {year}
  {2010})\BibitemShut {NoStop}%
\bibitem [{\citenamefont {Weidenm\"uller}\ and\ \citenamefont
  {Mitchell}(2009)}]{Weidemuller09}%
  \BibitemOpen
  \bibfield  {author} {\bibinfo {author} {\bibfnamefont {H.~A.}\ \bibnamefont
  {Weidenm\"uller}}\ and\ \bibinfo {author} {\bibfnamefont {G.~E.}\
  \bibnamefont {Mitchell}},\ }\bibfield  {title} {\bibinfo {title} {{Random
  matrices and chaos in nuclear physics: Nuclear structure}},\ }\href
  {https://doi.org/10.1103/RevModPhys.81.539} {\bibfield  {journal} {\bibinfo
  {journal} {Rev. Mod. Phys.}\ }\textbf {\bibinfo {volume} {81}},\ \bibinfo
  {pages} {539} (\bibinfo {year} {2009})}\BibitemShut {NoStop}%
\bibitem [{\citenamefont {Borgonovi}\ \emph {et~al.}(2016)\citenamefont
  {Borgonovi}, \citenamefont {Izrailev}, \citenamefont {Santos},\ and\
  \citenamefont {Zelevinsky}}]{Borgonovi16}%
  \BibitemOpen
  \bibfield  {author} {\bibinfo {author} {\bibfnamefont {F.}~\bibnamefont
  {Borgonovi}}, \bibinfo {author} {\bibfnamefont {F.~M.}\ \bibnamefont
  {Izrailev}}, \bibinfo {author} {\bibfnamefont {L.~F.}\ \bibnamefont
  {Santos}},\ and\ \bibinfo {author} {\bibfnamefont {V.~G.}\ \bibnamefont
  {Zelevinsky}},\ }\bibfield  {title} {\bibinfo {title} {Quantum chaos and
  thermalization in isolated systems of interacting particles},\ }\href
  {https://doi.org/10.1016/j.physrep.2016.02.005} {\bibfield  {journal}
  {\bibinfo  {journal} {Phys. Rep.}\ }\textbf {\bibinfo {volume} {626}},\
  \bibinfo {pages} {1} (\bibinfo {year} {2016})}\BibitemShut {NoStop}%
\bibitem [{\citenamefont {Beenakker}(1997)}]{Beenakker97}%
  \BibitemOpen
  \bibfield  {author} {\bibinfo {author} {\bibfnamefont {C.~W.~J.}\
  \bibnamefont {Beenakker}},\ }\bibfield  {title} {\bibinfo {title}
  {Random-matrix theory of quantum transport},\ }\href
  {https://doi.org/10.1103/RevModPhys.69.731} {\bibfield  {journal} {\bibinfo
  {journal} {Rev. Mod. Phys.}\ }\textbf {\bibinfo {volume} {69}},\ \bibinfo
  {pages} {731} (\bibinfo {year} {1997})}\BibitemShut {NoStop}%
\bibitem [{\citenamefont {Beenakker}(2015)}]{Beenakker15}%
  \BibitemOpen
  \bibfield  {author} {\bibinfo {author} {\bibfnamefont {C.~W.~J.}\
  \bibnamefont {Beenakker}},\ }\bibfield  {title} {\bibinfo {title}
  {{Random-matrix theory of Majorana fermions and topological
  superconductors}},\ }\href {https://doi.org/10.1103/RevModPhys.87.1037}
  {\bibfield  {journal} {\bibinfo  {journal} {Rev. Mod. Phys.}\ }\textbf
  {\bibinfo {volume} {87}},\ \bibinfo {pages} {1037} (\bibinfo {year}
  {2015})}\BibitemShut {NoStop}%
\bibitem [{\citenamefont {Maldacena}\ and\ \citenamefont
  {Stanford}(2016)}]{Maldacena16}%
  \BibitemOpen
  \bibfield  {author} {\bibinfo {author} {\bibfnamefont {J.}~\bibnamefont
  {Maldacena}}\ and\ \bibinfo {author} {\bibfnamefont {D.}~\bibnamefont
  {Stanford}},\ }\bibfield  {title} {\bibinfo {title} {{Remarks on the
  Sachdev-Ye-Kitaev model}},\ }\href
  {https://doi.org/10.1103/PhysRevD.94.106002} {\bibfield  {journal} {\bibinfo
  {journal} {Phys. Rev. D}\ }\textbf {\bibinfo {volume} {94}},\ \bibinfo
  {pages} {106002} (\bibinfo {year} {2016})}\BibitemShut {NoStop}%
\bibitem [{\citenamefont {Cotler}\ \emph {et~al.}(2017)\citenamefont {Cotler},
  \citenamefont {Gur-Ari}, \citenamefont {Hanada}, \citenamefont {Polchinski},
  \citenamefont {Saad}, \citenamefont {Shenker}, \citenamefont {Stanford},
  \citenamefont {Streicher},\ and\ \citenamefont {Tezuka}}]{Cotler18}%
  \BibitemOpen
  \bibfield  {author} {\bibinfo {author} {\bibfnamefont {J.~S.}\ \bibnamefont
  {Cotler}}, \bibinfo {author} {\bibfnamefont {G.}~\bibnamefont {Gur-Ari}},
  \bibinfo {author} {\bibfnamefont {M.}~\bibnamefont {Hanada}}, \bibinfo
  {author} {\bibfnamefont {J.}~\bibnamefont {Polchinski}}, \bibinfo {author}
  {\bibfnamefont {P.}~\bibnamefont {Saad}}, \bibinfo {author} {\bibfnamefont
  {S.~H.}\ \bibnamefont {Shenker}}, \bibinfo {author} {\bibfnamefont
  {D.}~\bibnamefont {Stanford}}, \bibinfo {author} {\bibfnamefont
  {A.}~\bibnamefont {Streicher}},\ and\ \bibinfo {author} {\bibfnamefont
  {M.}~\bibnamefont {Tezuka}},\ }\bibfield  {title} {\bibinfo {title} {Black
  holes and random matrices},\ }\href {https://doi.org/10.1007/JHEP05(2017)118}
  {\bibfield  {journal} {\bibinfo  {journal} {J. High Energy Phys.}\ }\textbf
  {\bibinfo {volume} {2017}},\ \bibinfo {pages} {118}}\BibitemShut {NoStop}%
\bibitem [{\citenamefont {D'Alessio}\ \emph {et~al.}(2016)\citenamefont
  {D'Alessio}, \citenamefont {Kafri}, \citenamefont {Polkovnikov},\ and\
  \citenamefont {Rigol}}]{DAlessio16}%
  \BibitemOpen
  \bibfield  {author} {\bibinfo {author} {\bibfnamefont {L.}~\bibnamefont
  {D'Alessio}}, \bibinfo {author} {\bibfnamefont {L.}~\bibnamefont {Kafri}},
  \bibinfo {author} {\bibfnamefont {A.}~\bibnamefont {Polkovnikov}},\ and\
  \bibinfo {author} {\bibfnamefont {M.}~\bibnamefont {Rigol}},\ }\bibfield
  {title} {\bibinfo {title} {From quantum chaos and eigenstate thermalization
  to statistical mechanics and thermodynamics},\ }\href
  {https://doi.org/10.1080/00018732.2016.1198134} {\bibfield  {journal}
  {\bibinfo  {journal} {Adv. Phys.}\ }\textbf {\bibinfo {volume} {65}},\
  \bibinfo {pages} {239} (\bibinfo {year} {2016})}\BibitemShut {NoStop}%
\bibitem [{\citenamefont {Deutsch}(2018)}]{Deutsch18}%
  \BibitemOpen
  \bibfield  {author} {\bibinfo {author} {\bibfnamefont {J.~M.}\ \bibnamefont
  {Deutsch}},\ }\bibfield  {title} {\bibinfo {title} {Eigenstate thermalization
  hypothesis},\ }\href {https://doi.org/10.1088/1361-6633/aac9f1} {\bibfield
  {journal} {\bibinfo  {journal} {Rep. Prog. Phys.}\ }\textbf {\bibinfo
  {volume} {81}},\ \bibinfo {pages} {082001} (\bibinfo {year}
  {2018})}\BibitemShut {NoStop}%
\bibitem [{\citenamefont {Kravtsv}\ \emph {et~al.}(2015)\citenamefont
  {Kravtsv}, \citenamefont {Khaymovich}, \citenamefont {Cuevas},\ and\
  \citenamefont {Amini}}]{Kravtsov15}%
  \BibitemOpen
  \bibfield  {author} {\bibinfo {author} {\bibfnamefont {V.~E.}\ \bibnamefont
  {Kravtsv}}, \bibinfo {author} {\bibfnamefont {I.~M.}\ \bibnamefont
  {Khaymovich}}, \bibinfo {author} {\bibfnamefont {E.}~\bibnamefont {Cuevas}},\
  and\ \bibinfo {author} {\bibfnamefont {M.}~\bibnamefont {Amini}},\ }\bibfield
   {title} {\bibinfo {title} {A random matrix model with localization and
  ergodic transitions},\ }\href
  {https://doi.org/10.1088/1367-2630/17/12/122002} {\bibfield  {journal}
  {\bibinfo  {journal} {New J. Phys.}\ }\textbf {\bibinfo {volume} {17}},\
  \bibinfo {pages} {122002} (\bibinfo {year} {2015})}\BibitemShut {NoStop}%
\bibitem [{\citenamefont {von Keyserlingk}\ \emph {et~al.}(2018)\citenamefont
  {von Keyserlingk}, \citenamefont {Rakovszky}, \citenamefont {Pollmann},\ and\
  \citenamefont {Sondhi}}]{vonKeyserlingk18}%
  \BibitemOpen
  \bibfield  {author} {\bibinfo {author} {\bibfnamefont {C.~W.}\ \bibnamefont
  {von Keyserlingk}}, \bibinfo {author} {\bibfnamefont {T.}~\bibnamefont
  {Rakovszky}}, \bibinfo {author} {\bibfnamefont {F.}~\bibnamefont
  {Pollmann}},\ and\ \bibinfo {author} {\bibfnamefont {S.~L.}\ \bibnamefont
  {Sondhi}},\ }\bibfield  {title} {\bibinfo {title} {{Operator Hydrodynamics,
  OTOCs, and Entanglement Growth in Systems without Conservation Laws}},\
  }\href {https://doi.org/10.1103/PhysRevX.8.021013} {\bibfield  {journal}
  {\bibinfo  {journal} {Phys. Rev. X}\ }\textbf {\bibinfo {volume} {8}},\
  \bibinfo {pages} {021013} (\bibinfo {year} {2018})}\BibitemShut {NoStop}%
\bibitem [{\citenamefont {Nahum}\ \emph {et~al.}(2018)\citenamefont {Nahum},
  \citenamefont {Vijay},\ and\ \citenamefont {Haah}}]{Nahum18}%
  \BibitemOpen
  \bibfield  {author} {\bibinfo {author} {\bibfnamefont {A.}~\bibnamefont
  {Nahum}}, \bibinfo {author} {\bibfnamefont {S.}~\bibnamefont {Vijay}},\ and\
  \bibinfo {author} {\bibfnamefont {J.}~\bibnamefont {Haah}},\ }\bibfield
  {title} {\bibinfo {title} {{Operator Spreading in Random Unitary Circuits}},\
  }\href {https://doi.org/10.1103/PhysRevX.8.021014} {\bibfield  {journal}
  {\bibinfo  {journal} {Phys. Rev. X}\ }\textbf {\bibinfo {volume} {8}},\
  \bibinfo {pages} {021014} (\bibinfo {year} {2018})}\BibitemShut {NoStop}%
\bibitem [{\citenamefont {Chan}\ \emph {et~al.}(2018)\citenamefont {Chan},
  \citenamefont {De~Luca},\ and\ \citenamefont {Chalker}}]{Chan18}%
  \BibitemOpen
  \bibfield  {author} {\bibinfo {author} {\bibfnamefont {A.}~\bibnamefont
  {Chan}}, \bibinfo {author} {\bibfnamefont {A.}~\bibnamefont {De~Luca}},\ and\
  \bibinfo {author} {\bibfnamefont {J.~T.}\ \bibnamefont {Chalker}},\
  }\bibfield  {title} {\bibinfo {title} {{Solution of a Minimal Model for
  Many-Body Quantum Chaos}},\ }\href
  {https://doi.org/10.1103/PhysRevX.8.041019} {\bibfield  {journal} {\bibinfo
  {journal} {Phys. Rev. X}\ }\textbf {\bibinfo {volume} {8}},\ \bibinfo {pages}
  {041019} (\bibinfo {year} {2018})}\BibitemShut {NoStop}%
\bibitem [{\citenamefont {Forrester}(2010)}]{Forrester10}%
  \BibitemOpen
  \bibfield  {author} {\bibinfo {author} {\bibfnamefont {P.~J.}\ \bibnamefont
  {Forrester}},\ }\href@noop {} {\emph {\bibinfo {title} {{Log-Gases and Random
  Matrices}}}}\ (\bibinfo  {publisher} {Princeton University Press},\ \bibinfo
  {address} {Princeton and Oxford},\ \bibinfo {year} {2010})\BibitemShut
  {NoStop}%
\bibitem [{\citenamefont {Livan}\ \emph {et~al.}(2018)\citenamefont {Livan},
  \citenamefont {Novaes},\ and\ \citenamefont {Vivo}}]{Livan18}%
  \BibitemOpen
  \bibfield  {author} {\bibinfo {author} {\bibfnamefont {G.}~\bibnamefont
  {Livan}}, \bibinfo {author} {\bibfnamefont {M.}~\bibnamefont {Novaes}},\ and\
  \bibinfo {author} {\bibfnamefont {P.}~\bibnamefont {Vivo}},\ }\href
  {https://doi.org/10.1007/978-3-319-70885-0} {\emph {\bibinfo {title}
  {{Introduction to Random Matrices: Theory and Practice}}}}\ (\bibinfo
  {publisher} {Springer},\ \bibinfo {address} {New York},\ \bibinfo {year}
  {2018})\BibitemShut {NoStop}%
\bibitem [{\citenamefont {Wachter}(1978)}]{Wachter78}%
  \BibitemOpen
  \bibfield  {author} {\bibinfo {author} {\bibfnamefont {K.~W.}\ \bibnamefont
  {Wachter}},\ }\bibfield  {title} {\bibinfo {title} {The strong limits of
  random matrix spectra for sample matrices of independent elements},\ }\href
  {https://doi.org/10.1214/aop/1176995607} {\bibfield  {journal} {\bibinfo
  {journal} {Ann. Prob.}\ }\textbf {\bibinfo {volume} {6}},\ \bibinfo {pages}
  {1} (\bibinfo {year} {1978})}\BibitemShut {NoStop}%
\bibitem [{\citenamefont {Venturelli}\ \emph {et~al.}(2022)\citenamefont
  {Venturelli}, \citenamefont {Cugliandolo}, \citenamefont {Schehr},\ and\
  \citenamefont {Tarzia}}]{Venturelli22}%
  \BibitemOpen
  \bibfield  {author} {\bibinfo {author} {\bibfnamefont {D.}~\bibnamefont
  {Venturelli}}, \bibinfo {author} {\bibfnamefont {L.~F.}\ \bibnamefont
  {Cugliandolo}}, \bibinfo {author} {\bibfnamefont {G.}~\bibnamefont
  {Schehr}},\ and\ \bibinfo {author} {\bibfnamefont {M.}~\bibnamefont
  {Tarzia}},\ }\bibfield  {title} {\bibinfo {title} {{Replica approach to the
  generalized Rosenzweig-Porter model}},\ }\Eprint
  {https://arxiv.org/abs/2209.11732} {arXiv:2209.11732}  (\bibinfo {year}
  {2022})\BibitemShut {NoStop}%
\bibitem [{\citenamefont {Bruder}\ and\ \citenamefont
  {Littlejohn}(2014)}]{Bruder14}%
  \BibitemOpen
  \bibfield  {author} {\bibinfo {author} {\bibfnamefont {A.}~\bibnamefont
  {Bruder}}\ and\ \bibinfo {author} {\bibfnamefont {L.~L.}\ \bibnamefont
  {Littlejohn}},\ }\bibfield  {title} {\bibinfo {title} {{Classical and Sobolev
  orthogonality of the nonclassical Jacobi polynomials with parameters $\alpha
  = \beta = -1$}},\ }\href {https://doi.org/10.1007/s10231-012-0284-8}
  {\bibfield  {journal} {\bibinfo  {journal} {Ann. Mat. Pura Appl.}\ }\textbf
  {\bibinfo {volume} {193}},\ \bibinfo {pages} {431} (\bibinfo {year}
  {2014})}\BibitemShut {NoStop}%
\bibitem [{\citenamefont {Sakhnovich}(2015)}]{Sakhnovich15}%
  \BibitemOpen
  \bibfield  {author} {\bibinfo {author} {\bibfnamefont {L.~A.}\ \bibnamefont
  {Sakhnovich}},\ }\href {https://doi.org/10.1007/978-3-319-16489-2} {\emph
  {\bibinfo {title} {{Integral Equations with Difference Kernels on Finite
  Intervals}}}},\ \bibinfo {edition} {2nd}\ ed.,\ \bibinfo {series} {{Operator
  Theory: Advances and Applications}}, Vol.~\bibinfo {volume} {84}\ (\bibinfo
  {publisher} {Birkh\"auser},\ \bibinfo {address} {Cham},\ \bibinfo {year}
  {2015})\BibitemShut {NoStop}%
\bibitem [{\citenamefont {Hardy}(2012)}]{Hardy12}%
  \BibitemOpen
  \bibfield  {author} {\bibinfo {author} {\bibfnamefont {A.}~\bibnamefont
  {Hardy}},\ }\bibfield  {title} {\bibinfo {title} {{A note on large deviations
  for 2D Coulomb gas with weakly confining potential}},\ }\href
  {https://doi.org/10.1214/ECP.v17-1818} {\bibfield  {journal} {\bibinfo
  {journal} {Electron. Commun. Probab.}\ }\textbf {\bibinfo {volume} {17}},\
  \bibinfo {pages} {1} (\bibinfo {year} {2012})}\BibitemShut {NoStop}%
\bibitem [{\citenamefont {Liflyand}(2021)}]{Liflyand21}%
  \BibitemOpen
  \bibfield  {author} {\bibinfo {author} {\bibfnamefont {E.}~\bibnamefont
  {Liflyand}},\ }\href {https://doi.org/10.1007/978-3-030-81892-0} {\emph
  {\bibinfo {title} {{Harmonic Analysis on the Real Line}}}}\ (\bibinfo
  {publisher} {Birkh\"auser},\ \bibinfo {address} {Cham},\ \bibinfo {year}
  {2021})\BibitemShut {NoStop}%
\bibitem [{\citenamefont {Akemann}\ \emph {et~al.}(2015)\citenamefont
  {Akemann}, \citenamefont {Baik},\ and\ \citenamefont
  {Di~Francesco}}]{Akemann15}%
  \BibitemOpen
  \bibinfo {editor} {\bibfnamefont {G.}~\bibnamefont {Akemann}}, \bibinfo
  {editor} {\bibfnamefont {J.}~\bibnamefont {Baik}},\ and\ \bibinfo {editor}
  {\bibfnamefont {P.}~\bibnamefont {Di~Francesco}},\ eds.,\ \href
  {https://doi.org/10.1093/oxfordhb/9780198744191.001.0001} {\emph {\bibinfo
  {title} {{The Oxford Handbook of Random Matrix Theory}}}}\ (\bibinfo
  {publisher} {Oxford University Press},\ \bibinfo {address} {Oxford},\
  \bibinfo {year} {2015})\BibitemShut {NoStop}%
\bibitem [{\citenamefont {Muttalib}\ \emph {et~al.}(1993)\citenamefont
  {Muttalib}, \citenamefont {Chen}, \citenamefont {Ismail},\ and\ \citenamefont
  {Nicopoulos}}]{Muttalib93}%
  \BibitemOpen
  \bibfield  {author} {\bibinfo {author} {\bibfnamefont {K.~A.}\ \bibnamefont
  {Muttalib}}, \bibinfo {author} {\bibfnamefont {Y.}~\bibnamefont {Chen}},
  \bibinfo {author} {\bibfnamefont {M.~E.~H.}\ \bibnamefont {Ismail}},\ and\
  \bibinfo {author} {\bibfnamefont {V.~N.}\ \bibnamefont {Nicopoulos}},\
  }\bibfield  {title} {\bibinfo {title} {{New Family of Unitary Random
  Matrices}},\ }\href {https://doi.org/10.1103/PhysRevLett.71.471} {\bibfield
  {journal} {\bibinfo  {journal} {Phys. Rev. Lett.}\ }\textbf {\bibinfo
  {volume} {71}},\ \bibinfo {pages} {471} (\bibinfo {year} {1993})}\BibitemShut
  {NoStop}%
\bibitem [{\citenamefont {Evers}\ and\ \citenamefont {Mirlin}(2008)}]{Evers08}%
  \BibitemOpen
  \bibfield  {author} {\bibinfo {author} {\bibfnamefont {F.}~\bibnamefont
  {Evers}}\ and\ \bibinfo {author} {\bibfnamefont {A.~D.}\ \bibnamefont
  {Mirlin}},\ }\bibfield  {title} {\bibinfo {title} {Anderson transitions},\
  }\href {https://doi.org/10.1103/RevModPhys.80.1355} {\bibfield  {journal}
  {\bibinfo  {journal} {Rev. Mod. Phys.}\ }\textbf {\bibinfo {volume} {80}},\
  \bibinfo {pages} {1355} (\bibinfo {year} {2008})}\BibitemShut {NoStop}%
\bibitem [{\citenamefont {Choi}\ and\ \citenamefont {Muttalib}(2010)}]{Choi10}%
  \BibitemOpen
  \bibfield  {author} {\bibinfo {author} {\bibfnamefont {J.}~\bibnamefont
  {Choi}}\ and\ \bibinfo {author} {\bibfnamefont {K.~A.}\ \bibnamefont
  {Muttalib}},\ }\bibfield  {title} {\bibinfo {title} {Universality of a family
  of random matrix ensembles with logarithmic soft-confinement potentials},\
  }\href {https://doi.org/10.1103/PhysRevB.82.104202} {\bibfield  {journal}
  {\bibinfo  {journal} {Phys. Rev. B}\ }\textbf {\bibinfo {volume} {82}},\
  \bibinfo {pages} {104202} (\bibinfo {year} {2010})}\BibitemShut {NoStop}%
\bibitem [{\citenamefont {Vleeshouwers}\ and\ \citenamefont
  {Gritsev}(2021)}]{Vleeshouwers21}%
  \BibitemOpen
  \bibfield  {author} {\bibinfo {author} {\bibfnamefont {W.~L.}\ \bibnamefont
  {Vleeshouwers}}\ and\ \bibinfo {author} {\bibfnamefont {V.}~\bibnamefont
  {Gritsev}},\ }\bibfield  {title} {\bibinfo {title} {{Topological field theory
  approach to intermediate statistics}},\ }\href
  {https://doi.org/10.21468/SciPostPhys.10.6.146} {\bibfield  {journal}
  {\bibinfo  {journal} {SciPost Phys.}\ }\textbf {\bibinfo {volume} {10}},\
  \bibinfo {pages} {146} (\bibinfo {year} {2021})}\BibitemShut {NoStop}%
\bibitem [{\citenamefont {Forrester}(2006)}]{Forrester06}%
  \BibitemOpen
  \bibfield  {author} {\bibinfo {author} {\bibfnamefont {P.~J.}\ \bibnamefont
  {Forrester}},\ }\bibfield  {title} {\bibinfo {title} {{Quantum conductance
  problems and the Jacobi ensemble}},\ }\href
  {https://doi.org/10.1088/0305-4470/39/22/004} {\bibfield  {journal} {\bibinfo
   {journal} {J. Phys. A: Math. Gen.}\ }\textbf {\bibinfo {volume} {39}},\
  \bibinfo {pages} {6861} (\bibinfo {year} {2006})}\BibitemShut {NoStop}%
\bibitem [{\citenamefont {Liu}\ \emph {et~al.}(2018)\citenamefont {Liu},
  \citenamefont {Chen},\ and\ \citenamefont {Balents}}]{Liu18}%
  \BibitemOpen
  \bibfield  {author} {\bibinfo {author} {\bibfnamefont {C.}~\bibnamefont
  {Liu}}, \bibinfo {author} {\bibfnamefont {X.}~\bibnamefont {Chen}},\ and\
  \bibinfo {author} {\bibfnamefont {L.}~\bibnamefont {Balents}},\ }\bibfield
  {title} {\bibinfo {title} {{Quantum entanglement of the Sachdev-Ye-Kitaev
  models}},\ }\href {https://doi.org/10.1103/PhysRevB.97.245126} {\bibfield
  {journal} {\bibinfo  {journal} {Phys. Rev. B}\ }\textbf {\bibinfo {volume}
  {97}},\ \bibinfo {pages} {245126} (\bibinfo {year} {2018})}\BibitemShut
  {NoStop}%
\bibitem [{\citenamefont {\L{}yd\.{z}ba}\ \emph {et~al.}(2021)\citenamefont
  {\L{}yd\.{z}ba}, \citenamefont {Rigol},\ and\ \citenamefont
  {Vidmar}}]{Lydzba21}%
  \BibitemOpen
  \bibfield  {author} {\bibinfo {author} {\bibfnamefont {P.}~\bibnamefont
  {\L{}yd\.{z}ba}}, \bibinfo {author} {\bibfnamefont {M.}~\bibnamefont
  {Rigol}},\ and\ \bibinfo {author} {\bibfnamefont {L.}~\bibnamefont
  {Vidmar}},\ }\bibfield  {title} {\bibinfo {title} {{Entanglement in many-body
  eigenstates of quantum-chaotic quadratic Hamiltonians}},\ }\href
  {https://doi.org/10.1103/PhysRevB.103.104206} {\bibfield  {journal} {\bibinfo
   {journal} {Phys. Rev. B}\ }\textbf {\bibinfo {volume} {103}},\ \bibinfo
  {pages} {104206} (\bibinfo {year} {2021})}\BibitemShut {NoStop}%
\bibitem [{\citenamefont {Bianchi}\ \emph {et~al.}(2021)\citenamefont
  {Bianchi}, \citenamefont {Hackl},\ and\ \citenamefont {Kieburg}}]{Bianchi21}%
  \BibitemOpen
  \bibfield  {author} {\bibinfo {author} {\bibfnamefont {E.}~\bibnamefont
  {Bianchi}}, \bibinfo {author} {\bibfnamefont {L.}~\bibnamefont {Hackl}},\
  and\ \bibinfo {author} {\bibfnamefont {M.}~\bibnamefont {Kieburg}},\
  }\bibfield  {title} {\bibinfo {title} {{Page curve for fermionic Gaussian
  states}},\ }\href {https://doi.org/10.1103/PhysRevB.103.L241118} {\bibfield
  {journal} {\bibinfo  {journal} {Phys. Rev. B}\ }\textbf {\bibinfo {volume}
  {103}},\ \bibinfo {pages} {L241118} (\bibinfo {year} {2021})}\BibitemShut
  {NoStop}%
\bibitem [{\citenamefont {Murciano}\ \emph {et~al.}(2022)\citenamefont
  {Murciano}, \citenamefont {Calabrese},\ and\ \citenamefont
  {Piroli}}]{Murciano22}%
  \BibitemOpen
  \bibfield  {author} {\bibinfo {author} {\bibfnamefont {S.}~\bibnamefont
  {Murciano}}, \bibinfo {author} {\bibfnamefont {P.}~\bibnamefont
  {Calabrese}},\ and\ \bibinfo {author} {\bibfnamefont {L.}~\bibnamefont
  {Piroli}},\ }\bibfield  {title} {\bibinfo {title} {{Symmetry-resolved Page
  curves}},\ }\href {https://doi.org/10.1103/PhysRevD.106.046015} {\bibfield
  {journal} {\bibinfo  {journal} {Phys. Rev. D}\ }\textbf {\bibinfo {volume}
  {106}},\ \bibinfo {pages} {046015} (\bibinfo {year} {2022})}\BibitemShut
  {NoStop}%
\bibitem [{\citenamefont {Ul\v{c}akar}\ and\ \citenamefont
  {Vidmar}(2022)}]{Ulcakar22}%
  \BibitemOpen
  \bibfield  {author} {\bibinfo {author} {\bibfnamefont {I.}~\bibnamefont
  {Ul\v{c}akar}}\ and\ \bibinfo {author} {\bibfnamefont {L.}~\bibnamefont
  {Vidmar}},\ }\bibfield  {title} {\bibinfo {title} {Tight-binding billiards},\
  }\href {https://doi.org/10.1103/PhysRevE.106.034118} {\bibfield  {journal}
  {\bibinfo  {journal} {Phys. Rev. E}\ }\textbf {\bibinfo {volume} {106}},\
  \bibinfo {pages} {034118} (\bibinfo {year} {2022})}\BibitemShut {NoStop}%
\bibitem [{\citenamefont {Flack}\ \emph {et~al.}(2020)\citenamefont {Flack},
  \citenamefont {Bertini},\ and\ \citenamefont {Prosen}}]{Flack20}%
  \BibitemOpen
  \bibfield  {author} {\bibinfo {author} {\bibfnamefont {A.}~\bibnamefont
  {Flack}}, \bibinfo {author} {\bibfnamefont {B.}~\bibnamefont {Bertini}},\
  and\ \bibinfo {author} {\bibfnamefont {T.}~\bibnamefont {Prosen}},\
  }\bibfield  {title} {\bibinfo {title} {{Statistics of the spectral form
  factor in the self-dual kicked Ising model}},\ }\href
  {https://doi.org/10.1103/PhysRevResearch.2.043403} {\bibfield  {journal}
  {\bibinfo  {journal} {Phys. Rev. Research}\ }\textbf {\bibinfo {volume}
  {2}},\ \bibinfo {pages} {043403} (\bibinfo {year} {2020})}\BibitemShut
  {NoStop}%
\bibitem [{\citenamefont {Koekoek}\ \emph {et~al.}(2010)\citenamefont
  {Koekoek}, \citenamefont {Lesky},\ and\ \citenamefont
  {Swarttouw}}]{Koekoek10}%
  \BibitemOpen
  \bibfield  {author} {\bibinfo {author} {\bibfnamefont {R.}~\bibnamefont
  {Koekoek}}, \bibinfo {author} {\bibfnamefont {P.~A.}\ \bibnamefont {Lesky}},\
  and\ \bibinfo {author} {\bibfnamefont {R.~F.}\ \bibnamefont {Swarttouw}},\
  }\href {https://doi.org/10.1007/978-3-642-05014-5} {\emph {\bibinfo {title}
  {{Hypergeometric Orthogonal Polynomials and Their $q$-Analogues}}}}\
  (\bibinfo  {publisher} {Springer},\ \bibinfo {address} {New York},\ \bibinfo
  {year} {2010})\BibitemShut {NoStop}%
\bibitem [{\citenamefont {Szeg\"{o}}(1975)}]{Szego75}%
  \BibitemOpen
  \bibfield  {author} {\bibinfo {author} {\bibfnamefont {G.}~\bibnamefont
  {Szeg\"{o}}},\ }\href@noop {} {\emph {\bibinfo {title} {{Orthogonal
  Polynomials}}}},\ \bibinfo {edition} {4th}\ ed.,\ \bibinfo {series}
  {{Colloquium Publications}}, Vol.~\bibinfo {volume} {23}\ (\bibinfo
  {publisher} {American Mathematical Society},\ \bibinfo {address} {Providence,
  Rhode Island},\ \bibinfo {year} {1975})\BibitemShut {NoStop}%
\bibitem [{\citenamefont {Moreno-Pozas}\ \emph {et~al.}(2019)\citenamefont
  {Moreno-Pozas}, \citenamefont {Morales-Jimenez},\ and\ \citenamefont
  {McKay}}]{MorenoPozas19}%
  \BibitemOpen
  \bibfield  {author} {\bibinfo {author} {\bibfnamefont {L.}~\bibnamefont
  {Moreno-Pozas}}, \bibinfo {author} {\bibfnamefont {D.}~\bibnamefont
  {Morales-Jimenez}},\ and\ \bibinfo {author} {\bibfnamefont {M.~R.}\
  \bibnamefont {McKay}},\ }\bibfield  {title} {\bibinfo {title} {{Extreme
  eigenvalue distributions of Jacobi ensembles: New exact representations,
  asymptotics and finite size corrections}},\ }\href
  {https://doi.org/10.1016/j.nuclphysb.2019.114724} {\bibfield  {journal}
  {\bibinfo  {journal} {Nucl. Phys. B}\ }\textbf {\bibinfo {volume} {947}},\
  \bibinfo {pages} {114724} (\bibinfo {year} {2019})}\BibitemShut {NoStop}%
\bibitem [{\citenamefont {Dumitriu}\ and\ \citenamefont
  {Edelman}(2002)}]{Dumitriu02}%
  \BibitemOpen
  \bibfield  {author} {\bibinfo {author} {\bibfnamefont {I.}~\bibnamefont
  {Dumitriu}}\ and\ \bibinfo {author} {\bibfnamefont {A.}~\bibnamefont
  {Edelman}},\ }\bibfield  {title} {\bibinfo {title} {Matrix models for beta
  ensembles},\ }\href {https://doi.org/10.1063/1.1507823} {\bibfield  {journal}
  {\bibinfo  {journal} {J. Math. Phys.}\ }\textbf {\bibinfo {volume} {43}},\
  \bibinfo {pages} {5830} (\bibinfo {year} {2002})}\BibitemShut {NoStop}%
\bibitem [{\citenamefont {Khaymovich}\ \emph {et~al.}(2020)\citenamefont
  {Khaymovich}, \citenamefont {Kravtsov}, \citenamefont {Altshuler},\ and\
  \citenamefont {Ioffe}}]{Khaymovich20}%
  \BibitemOpen
  \bibfield  {author} {\bibinfo {author} {\bibfnamefont {I.~M.}\ \bibnamefont
  {Khaymovich}}, \bibinfo {author} {\bibfnamefont {V.~E.}\ \bibnamefont
  {Kravtsov}}, \bibinfo {author} {\bibfnamefont {B.~L.}\ \bibnamefont
  {Altshuler}},\ and\ \bibinfo {author} {\bibfnamefont {L.~B.}\ \bibnamefont
  {Ioffe}},\ }\bibfield  {title} {\bibinfo {title} {{Fragile extended phases in
  the log-normal Rosenzweig-Porter model}},\ }\href
  {https://doi.org/10.1103/PhysRevResearch.2.043346} {\bibfield  {journal}
  {\bibinfo  {journal} {Phys. Rev. Research}\ }\textbf {\bibinfo {volume}
  {2}},\ \bibinfo {pages} {043346} (\bibinfo {year} {2020})}\BibitemShut
  {NoStop}%
\bibitem [{\citenamefont {Biroli}\ and\ \citenamefont
  {Tarzia}(2021)}]{Biroli21}%
  \BibitemOpen
  \bibfield  {author} {\bibinfo {author} {\bibfnamefont {G.}~\bibnamefont
  {Biroli}}\ and\ \bibinfo {author} {\bibfnamefont {M.}~\bibnamefont
  {Tarzia}},\ }\bibfield  {title} {\bibinfo {title} {{L\'evy-Rosenzweig-Porter
  random matrix ensemble}},\ }\href
  {https://doi.org/10.1103/PhysRevB.103.104205} {\bibfield  {journal} {\bibinfo
   {journal} {Phys. Rev. B}\ }\textbf {\bibinfo {volume} {103}},\ \bibinfo
  {pages} {104205} (\bibinfo {year} {2021})}\BibitemShut {NoStop}%
\bibitem [{\citenamefont {Khaymovich}\ and\ \citenamefont
  {Kravtsov}(2021)}]{Khaymovich21}%
  \BibitemOpen
  \bibfield  {author} {\bibinfo {author} {\bibfnamefont {I.~M.}\ \bibnamefont
  {Khaymovich}}\ and\ \bibinfo {author} {\bibfnamefont {V.~E.}\ \bibnamefont
  {Kravtsov}},\ }\bibfield  {title} {\bibinfo {title} {{Dynamical phases in a
  ``multifractal'' Rosenzweig-Porter model}},\ }\href
  {https://doi.org/10.21468/SciPostPhys.11.2.045} {\bibfield  {journal}
  {\bibinfo  {journal} {SciPost Phys.}\ }\textbf {\bibinfo {volume} {11}},\
  \bibinfo {pages} {045} (\bibinfo {year} {2021})}\BibitemShut {NoStop}%
\bibitem [{\citenamefont {Buijsman}\ and\ \citenamefont
  {Bar~Lev}(2022)}]{Buijsman22}%
  \BibitemOpen
  \bibfield  {author} {\bibinfo {author} {\bibfnamefont {W.}~\bibnamefont
  {Buijsman}}\ and\ \bibinfo {author} {\bibfnamefont {Y.}~\bibnamefont
  {Bar~Lev}},\ }\bibfield  {title} {\bibinfo {title} {{Circular
  Rosenzweig-Porter random matrix ensemble}},\ }\href
  {https://doi.org/10.21468/SciPostPhys.12.3.082} {\bibfield  {journal}
  {\bibinfo  {journal} {SciPost Phys.}\ }\textbf {\bibinfo {volume} {12}},\
  \bibinfo {pages} {082} (\bibinfo {year} {2022})}\BibitemShut {NoStop}%
\bibitem [{\citenamefont {De~Tomasi}\ and\ \citenamefont
  {Khaymovich}(2022)}]{DeTomasi22}%
  \BibitemOpen
  \bibfield  {author} {\bibinfo {author} {\bibfnamefont {G.}~\bibnamefont
  {De~Tomasi}}\ and\ \bibinfo {author} {\bibfnamefont {I.~M.}\ \bibnamefont
  {Khaymovich}},\ }\bibfield  {title} {\bibinfo {title} {{Non-Hermitian
  Rosenzweig-Porter random-matrix ensemble: Obstruction to the fractal
  phase}},\ }\href {https://doi.org/10.1103/PhysRevB.106.094204} {\bibfield
  {journal} {\bibinfo  {journal} {Phys. Rev. B}\ }\textbf {\bibinfo {volume}
  {106}},\ \bibinfo {pages} {094204} (\bibinfo {year} {2022})}\BibitemShut
  {NoStop}%
\bibitem [{\citenamefont {Akemann}\ \emph {et~al.}(2021)\citenamefont
  {Akemann}, \citenamefont {Nagao}, \citenamefont {Parra},\ and\ \citenamefont
  {Vernizzi}}]{Akemann21}%
  \BibitemOpen
  \bibfield  {author} {\bibinfo {author} {\bibfnamefont {G.}~\bibnamefont
  {Akemann}}, \bibinfo {author} {\bibfnamefont {T.}~\bibnamefont {Nagao}},
  \bibinfo {author} {\bibfnamefont {I.}~\bibnamefont {Parra}},\ and\ \bibinfo
  {author} {\bibfnamefont {G.}~\bibnamefont {Vernizzi}},\ }\bibfield  {title}
  {\bibinfo {title} {{Gegenbauer and Other Planar Orthogonal Polynomials on an
  Ellipse in the Complex Plane}},\ }\href
  {https://doi.org/10.1007/s00365-020-09515-0} {\bibfield  {journal} {\bibinfo
  {journal} {Constr. Approx.}\ }\textbf {\bibinfo {volume} {53}},\ \bibinfo
  {pages} {441} (\bibinfo {year} {2021})}\BibitemShut {NoStop}%
\end{thebibliography}%

\end{document}